\documentclass{ws-ijmpb}

\begin{document}

\markboth{Ning Ding, Yougui Wang, Jun Xu \and Ning Xi} {Power Law
Distribution in Circulating Money: Effect of Preferential
Behavior}

%
\catchline{}{}{}{}{}
%

\title{POWER-LAW DISTRIBUTIONS IN CIRCULATING MONEY: EFFECT OF PREFERENTIAL BEHAVIOR
}

\author{NING DING, YOUGUI WANG\footnote{Corresponding Author,
 E-mail: ygwang@bnu.edu.cn}, JUN XU \and NING XI
}

\address{Department of Systems Science, School of Management, Beijing Normal University, Beijing,
100875, People's Republic of China \\
}

\maketitle

\begin{history}
\received{Day Month Year}
\revised{Day Month Year}
\end{history}

\begin{abstract}
We introduce preferential behavior into the study on statistical
mechanics of money circulation. The computer simulation results
show that the preferential behavior can lead to power laws on
distributions over both holding time and amount of money held by
agents. However, some constraints are needed in generation
mechanism to ensure the robustness of power-law distributions.
\end{abstract}


\section{Introduction}

Empirical studies on many social issues show that numerous
statistical distributions follow power laws, such as stock-price
fluctuations\cite{stanley}, the probability distribution of the
population of cities\cite{susanna}, the degree distributions in
many networks\cite{barabasi} and distribution of income or
wealth\cite{pareto,empirical}. In particular, the wealth
distribution which is an old topic has recently been renewed by a
small band of econophysicists\cite{follow}. They proposed
multi-agent interaction models to explore the mechanism of wealth
distribution, in which wealth is simply represented by
money\cite{basic,saving}. However, to our knowledge, the
simulation results of these works show that the steady states of
money have Gibbs or Gamma distributions, leaving the power-law
phenomena of wealth distribution unexplained. It has been shown
that preferential behavior assumption plays an essential role in
generating power-law distributions in some cases, such as in
networks\cite{barabasi} and in money dynamics\cite{ding}. The soul
of preferential assumption is to break the equality among agents,
some of which are entitled to more power in getting corresponding
entity. In this paper, we investigate how the individual
preferential behavior generates power law phenomena in money
circulation process. The statistical distributions involved in
monetary circulation are composed of two aspects. One is over the
time interval the money stays in agents' hands which is named as
holding time\cite{wang}, the other is over the amount of money
held by agents.

\section{Power Law Distribution over Holding Time}

The model in this section is an extension of ideal gas-like
model\cite{basic}, where each agent is identified as a gas
molecule and each trading as one elastic (two-body) collision. In
this model, the economic system is assumed to be closed, thus the
total amount of money $M$ before and after transaction is
conserved and the number of agents $N$ remains constant. Since the
scale and initial distribution of money have no effect on the
final results, most of our simulations were carried out with
$N=250$ and $M=25,000$ and the amount of money held by each agent
was set to be $M/N$ at the beginning. The money is possessed by
agents individually and agents can exchange money with each other.
In each round, an arbitrary pair of agents i and j gets engaged in
a trade among which agent $i$ is randomly picked to be the
``payer'' and the other one $j$ becomes the ``receiver''. The
amount of money transferred is determined by the following
equation
\begin{equation}\label{deltam}
    \Delta m=\varepsilon(m_i+m_j)/2,
\end{equation}
where $\varepsilon$ distributes randomly and uniformly within an
interval $[0,1]$.

In the original ideal gas-like models, the money paid out is
chosen with equal probability, which has been discussed in
Ref.~\refcite{wang}. In the extension model, the preferential
behavior is introduced by imposing the unequal probability. In a
given round, agent $i$ with money $m_i$ in hand is the payer, the
probability of money $k$ among $m_i$ to be transferred is given
by:
\begin{equation}\label{pre}
    p(k)=\frac{l_k+1}{\sum_{n=1}^{m_i}(l_n+1)};
\end{equation}
where $l_n$ is the times that money $n$ has participated in trade
since the beginning of simulation. Here, we express the
probability with exchange times plus 1 instead of exchange times
in case that denominator be zero at the beginning of simulations.

In simulations, the interval between the first two exchanges for
one unit of money to participate in is recorded as its holding
time after most of money($\geq99.9\%$ in our simulations) had been
exchanged at least one time. Then, after most of money($\geq95\%$)
are recorded, the sampling of the holding times of money in this
system is completed.

\begin{figure}[th]
\begin{minipage}[t]{0.48\textwidth}
\centerline{\psfig{file=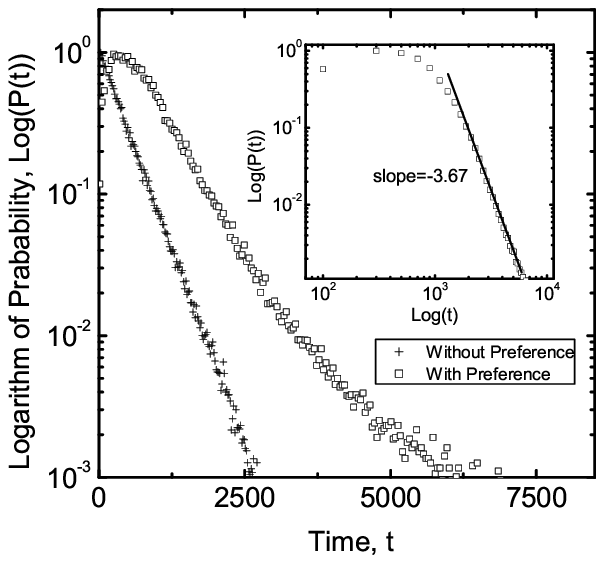,width=7 cm
}} \caption{The
stationary distributions of holding time with and without
preference in the semi-logarithmic scale. The inset shows the
distribution with preference in a double-logarithmic plot. Note
that in the figure the probabilities have been scaled by the
maximum probability.} \label{fig:side:a}
\end{minipage}
\ \
\begin{minipage}[t]{0.48\textwidth}
\centerline{\psfig{file=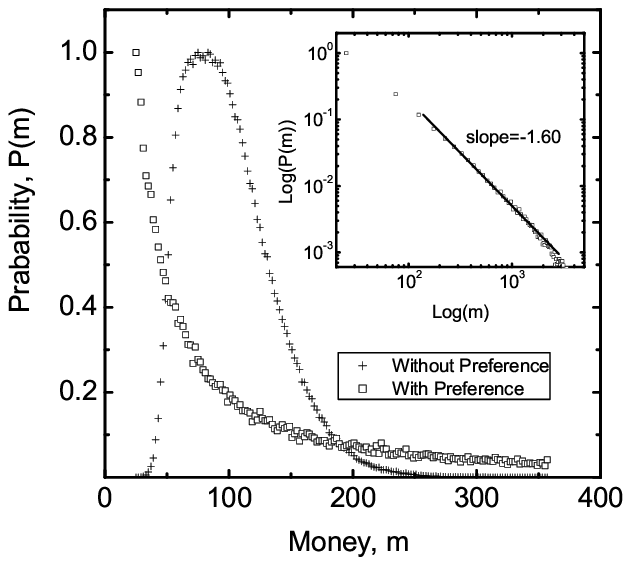,width=7 cm
}} \caption{The
stationary distributions of money with and without preference. The
inset shows the distribution with preference in a
double-logarithmic plot. Note that in the figure the probabilities
have been scaled by the maximum probability.} \label{fig:side:b}
\end{minipage}
\end{figure}

Each of the results shown in Figure 1 is an average of 500
simulations. It can be found that the holding time distribution
does not follow Gibbs-Boltzmann distribution any more after the
introduction of preferential behavior. There is an initial growth
of $P(t)$ from $t=0$, which quickly saturates and then a long
range of power-law decay in $P(t)$ for large $t$ value is
observed. This decay, when fitted to power law $P(t)\propto
t^{-\nu}$, gives $\nu=-3.67\pm 0.05$. We also examined the holding
time distributions in several periods long enough after the system
had achieved stationary state, and we found the power distribution
is remarkably robust, e.g., the holding time distribution is
stationary even after $t=500,000$ while it has been observed after
$t=1,000$.

\section{Power Law Distribution over Money Held}

Previous studies on money distribution are all carried out within
the framework of ideal gas-like models. We introduced the
preferential behavior into such kinds of model by assuming that
the agent with more money has larger probability to win or to be
chosen to participate in the trade. However slightly the
preferential propensity is set, we get the same final result: one
agent achieves all of the money. This indicates that preferential
behavior is not enough to produce robust power-law distributions
in such case. Thus, in this section, the effects of preferential
behavior on money distribution will be analyzed within a new
framework.

In what follows, the initial setting of the system is the same
with the ideal gas-like model, $N=250$, $M=25,000$ and each agent
has 100 units of money in hand at the beginning. The main novel
mechanism introduced here is to assume that every agent pays money
out to others in each round, and the amount of money paid out is
determined randomly. As to how to dispense the money, there are
two modes. One is that the others have equal probability to
receive each unit of money; the other one is that the probability
$p_{i,j}$ at which agent i gets the money from agent j satisfies
\begin{equation}\label{ptg}
    p_{i,j}=\frac{m_i}{\sum_{n=1 (n\neq j)}^{N}m_n},
\end{equation}
where $m_i$ is the amount of money held by agent $i$ before the
trade. Please note the constraint $n\neq j$ in this rule
eliminates the possibility for the payer to get back the money he
has paid out. It is obvious that the second mode is with
preference, in which the rich have higher probability to get
richer.

The simulation results of the two modes are shown in figure 2,
both of which are averages of 500 simulations. And they reveal
clearly that the preferential behavior does have effects on the
probability distribution of money. The stationary distribution of
money without preference is a Gamma type. After the preferential
behavior is introduced into the model, the power-law distribution
is observed, and the fitting to power law gives exponent
$\gamma=-1.60 \pm 0.02$. Further measurement performed after
$t=500,000$ shows that the distribution is quite robust.

In the simulations, if we removed the constraint $n\neq j$ in
Eq.(\ref{ptg}), we found that for $t=500,000$, more than 80\% of
the money was acquired by one agent. It can be forecasted that
after enough long time, all of money would be held by one agent.
This fact means that only preferential behavior without any
constraints can not induce power laws.

\section{Conclusion}
In this paper we studied the effect of preferential behavior on
probability distributions of both aspects of the circulation of
money. We performed computer simulations to show how the
preferential behavior produces power-law distributions over
holding time and money held respectively. It is also worth noting
that some constraints may be necessary to ensure the robustness of
power laws. The conclusion may be valuable to the understanding on
the mechanism of power laws.

\section*{Acknowledgements}
We would like to thank Prof. Zengru Di for encouragement and
enlightening communications. This work was supported by the the
National Science Foundation of China under Grant No. 70371072.


\begin{thebibliography}{0}
\bibitem{stanley} P. Bak, K. Chen, J.A. Scheinkman and M.
Woodford, {\it Ric. Economichi} {\bf 47}, 3 (1993); M.H.R. Stanley
{\it et al.} {\it Nature} {\bf 379}, 804 (1996).

\bibitem{susanna} S.C. Manrubia, D.H. Zanette, {\it Phys. Rev. E.}
{\bf 58}, 295 (1998).

\bibitem{barabasi} R. Albert and A.-L. Barab\'{a}si,{\it Reviews of Modern Physics} {\bf 74}, 47 (2002).

\bibitem{pareto} V. Pareto, {\it Cours d'Economie Politique\/}, (Librairie
Droz, Geneva, 1897).

\bibitem{empirical} A. Dr\u agulescu, and V. M. Yakovenko, {\it Physica A} \textbf{299}, 213 (2001);
A. Dr\u{a}gulescu and V. M. Yakovenko, {\it Eur. Phys. J. B}
\textbf{20} 585-589 (2001).

\bibitem{follow} B. Hayes, {\it Am. Scientist} {\bf90}, 400 (2002).

\bibitem{basic} A. A. Dr\u{a}gulescu and V. M. Yakovenko, {\it Eur. Phys. J. B} {\bf 17}, 723(2000).

\bibitem{saving} A. Chakraborti and B. K. Chakrabarti, {\it Eur.Phys. J. B} {\bf 17}, 167 (2000);
M. Patriarca, A. Chakraborti and K. Kaski, arXiv:
cond-mat/0312167.

\bibitem{ding} N. Ding, Y. Wang and Li Zhang, {\it Eur.Phys. J. B} {\bf 36}, 149-153(2003).



\bibitem{wang} Y. Wang, N. Ding and L. Zhang, {\it Physica A} {\bf 324}, 665-667(2003).






\end{thebibliography}
\end{document}